\documentstyle[twocolumn,epsf,prl,aps]{revtex}

\begin{document}

\draft

\title{ Vortex phase diagram in trapped Bose-Einstein condensation}

\author{Takeshi Mizushima, Tomoya Isoshima and Kazushige Machida}
\address{Department of Physics, Okayama University,
         Okayama 700-8530, Japan}
 \date{\today}

\maketitle

\begin{abstract}
The vortex phase diagram in the external rotation frequency
versus temperature is calculated 
for dilute Bose-Einstein condensed gases.
It is determined within the Bogoliubov-Popov theory for
a finite temperature where the condensate and non-condensate
fractions are treated in an equal footing.
The temperature dependences of various thermodynamic instability
lines for the vortex nucleation are computed to construct
the phase diagram.
Experiments are proposed to resolve a recent controversy
on the vortex creation problem associated with the quantized 
vortex observation in $^{87}$Rb atom gases. 
\end{abstract}


\section{Introduction}
Since its realization in 1995\cite{cornell,hulet,ketterle}, Bose-Einstein
condensation phenomena
(BEC) in alkali atom vapors  (Na, Li and Rb) have been attracting much
attention.
Several outstanding fundamental questions such as the relationship
between the origin of superfluidity and Bose-Einstein condensation can
now address directly both experimentally and
theoretically\cite{review1,review2}.
Recently long-sought quantized vortex has been successfully created in
$^{87}$Rb atom gases by two different experimental
methods\cite{matthews,anderson,madison1,chevy,madison2};
One is the so-called phase imprinting\cite{matthews,anderson} and the other
is to use
an optical spoon\cite{madison1,chevy,madison2}. The latter is conceptually
similar to the rotating
bucket experiment in
$^4$He and  $^3$He to create quantized vortices\cite{donnelly};
Here the harmonically trapped BEC with an elongated cigar shape is rotated
around the
long axis by stirring two
laser beams. As the stirred rotation angular frequency $\Omega$ increases,
a quantized vortex appears at the nucleation frequency $\Omega_{nuc}$, followed
by multiple vortices with each carrying the quantized unit circulation.
They form an Abrikosov-like lattice structure.
The resulting nucleation frequency $\Omega_{nuc}$ relative to the radial
confining
potential $\omega_{\perp}$ is $\Omega_{nuc}\sim0.68\omega_{\perp}$
which is not much dependent
on their experimental parameters such as the atom number, or the confining
potential, etc. in their experiments\cite{madison1,chevy,madison2}.
This particular number $\Omega_{nuc}\sim0.68\omega_{\perp}$
sharply disagrees with the prediction\cite{lundh} by the thermodynamic
critical frequency
$\Omega^L_{global}$ at which the single vortex state becomes stable
thermodynamically, namely, a theory based on Thomas-Fermi
approximation gives $\Omega^L_{global}/\omega_{\perp}\sim0.37$.
It turns out that the nucleation frequency happens to coincide with
the characteristic frequency $\Omega_{w=0}$ at which the non-vortex state
ultimately
 becomes unstable, leaving only the single vortex state stable globally in
 the energy configuration space. However, it is argued\cite{stringari1,sinha}
 that since the actual nucleation
 process is more or less related dynamical instability rather than the above
thermodymanic
 quantities, that is, the process is initiated by resonantly exciting the
quadra-pole mode
 in a slightly radially distorted cigar-shaped BEC.
 Thus there is no definite answer to explain $\Omega_{nuc}$ at present.

 The purpose of the present paper is to complete the vortex phase diagram in
the
 plane; the external rotation frequency $\Omega$ versus temperature $T$
 in order to help establishing vortex properties in general in the present
 dilute Bose gas systems, in particular, the nucleation
process by
 identifying the following characteristic critical
 frequencies related to various thermodynamic instabilities, 
 which will be found to have different
 $T$-dependence.

Previously in our series of
papers\cite{isoshima1,isoshima2,isoshima3,isoshima4,isoshima5},
we have identified several characteristic critical frequencies;
Upon increasing $\Omega$ from zero, the single vortex state first becomes
locally
stable at $\Omega^L_{local}$ and then stable globally at $\Omega^L_{global}$
relative to the non-vortex state. At further higher frequency
$\Omega_{w=0}$ the
non-vortex state exhibits an instability toward the vortex state and
finally the single vortex state becomes unstable, signalling that multiple
vortex state
can be stabilized. Here we will investigate the temperature evolution of
these critical
frequencies by
solving the Bogoliubov-de Gennes equation for non-condensates.
It is coupled
with the generalized Gross-Pitaevskii equation for the condensates
within the so-called Popov approximation.
The condensate and
non-condesates are treated in an equal footing.
We assume a cylindrical symmetric system confined radially by a harmonic
potential. The rotation occurs around the cylindrical axis.

Stringari\cite{stringari2} gives a phase diagram in the plane; $\Omega$ 
versus
$T$ and determines the $T$-dependence
of $\Omega^L_{global}$ in our terminology, 
basing on the Gross-Pitaevskii equation by neglecting
the non-condensate contribution. In contrast, we take into account thermally
excited non-condensate contribution fully in our calculation as mentioned.

This is one of our continuous efforts to understand the vortex matter in BEC.
We have first pointed out the presence of the anomalous negative mode
with the particular angular momentum both for the bucket-like system
with a rigid wall boundary\cite{isoshima1} and for the system
with the confining harmonic potential\cite{isoshima2}.
It turned out later that this anomalous mode in BEC is crucial to
understand the vortex matter, in particular,  the nucleation problem.
Since we find that the non-condensate contribution due to
finite temperature effect helps stabilizing the vortex state by pushing up
the negative mode, it is of interesting to know the vortex phase diagram
in $\Omega$ versus $T$.

The arrangement of the paper is as follows: The formulation  of the problem
is given within the Hartree-Fock Bogoliubov framework in next section.
Based on the self-consistent calculations the local stability of a vortex
is considered in III and the global stability is discussed on IV. The final
section is
devoted to constructing the phase diagram in $\Omega$ vs $T$ and discussions.

\section{Formulation}
The Hamiltonian in a rotating frame with the angular frequency with
$\omega$ is given by

\begin{eqnarray}
 \hat{H}_{rot} & = & \hat{H} -
          {\vec\omega} \cdot \int \hat{\Psi}^{\dagger} ( {\bf r} ) \cdot
           ( {\bf r} \times {\bf p} ) \cdot
                 \hat{\Psi} ( {\bf r} ) d {\bf r}  \\
 \hat{H} & = & \int 
             d {\bf r} \hat{\Psi}^{\dagger} ( {\bf r} )
             \{ \; - \frac{\hbar^2}{2m} \nabla^2 + V ( {\bf r} ) - \mu \; \}
             \hat{\Psi} ( {\bf r} ) \nonumber \\
           & + & \frac{g}{2} \int d {\bf r}
             \hat{\Psi}^{\dagger} ( {\bf r} ) \hat{\Psi}^{\dagger} ( {\bf r} )
             \hat{\Psi} ( {\bf r} ) \hat{\Psi} ( {\bf r} )
\end{eqnarray}

\noindent
where the creation and annihilation operators of the Bose particle
are $\Psi^{\dagger}( {\bf r} )$ and $\Psi ( {\bf r})$. They interact
each other with the contact interaction $g\equiv {4\pi \hbar^2 a\over m}$
($a$: the scattering length and $m$: the mass).
The confining potential $V( {\bf r} )$ and $\mu$ is the chemical potential.
These are decomposed into
$\hat{\Psi} ( {\bf r} ) = \phi ( {\bf r} ) + \hat{\psi} ( {\bf r} )$
in terms of the condensate wave function: $\phi ( {\bf r} )$ and the
non-condensate
part: $\hat{\psi} ( {\bf r} )$.
By this decomposition, the above Hamiltonian is rewritten as

\begin{eqnarray}
 \hat{H}_{rot} 
   & = & \int d {\bf r} \;
       [ \; \{ \; \phi^{\ast} ( {\bf r} ) h ( {\bf r} ) \phi ( {\bf r} )
       + \frac{1}{2} g { | \phi ( {\bf r} ) | }^4 \; \}  \nonumber \\
   & + & \hat{\psi}^{\dagger} ( {\bf r} )
       \{ \; h ( {\bf r} ) + g | \phi ( {\bf r} ) | ^2 \; \}
      \phi ( {\bf r} ) + h.c. \nonumber \\
   & + & \hat{\psi}^{\dagger} ( {\bf r} )
       \{ \; h ( {\bf r} ) + 2g | \phi ( {\bf r} ) | ^2 \; \}
      \hat{\psi} ( {\bf r} ) \nonumber \\
   & + & \frac{g}{2}
      \hat{\psi}^{\dagger} ( {\bf r} ) \hat{\psi}^{\dagger} ( {\bf r} )
      \phi ( {\bf r} ) \phi ( {\bf r} ) + h.c. \nonumber \\
   & + & g \hat{\psi}^{\dagger} ( {\bf r} ) \hat{\psi} ( {\bf r} )
      \hat{\psi} ( {\bf r} ) \phi^{\ast} ( {\bf r} ) + h.c. \nonumber \\
   & + & \frac{g}{2}
      \hat{\psi}^{\dagger} ( {\bf r} ) \hat{\psi}^{\dagger} ( {\bf r} )
      \hat{\psi} ( {\bf r} ) \hat{\psi} ( {\bf r} ) \nonumber \\
   & + & i \hbar
      \{ \; \phi^{\ast} ( {\bf r} ) {\vec \omega} \cdot ( {\bf r} \times
         \nabla ) \phi ( {\bf r} ) \nonumber \\
   & & + \hat{\psi}^{\dagger} ( {\bf r} ) {\vec \omega} \cdot ( {\bf r} 
         \times \nabla ) \phi ( {\bf r} ) \nonumber \\
   & & + \phi^{\ast} ( {\bf r} ) {\vec \omega} \cdot ( {\bf r} \times
         \nabla ) \hat{\psi} ( {\bf r} ) \nonumber \\
   & & + \hat{\psi}^{\dagger} ( {\bf r} ) {\vec \omega} \cdot ( {\bf r} 
         \times \nabla ) \hat{\psi} ( {\bf r} )
      \; \} \; ].
\end{eqnarray}

\noindent
The single particle Hamiltonian is
\begin{eqnarray}
 h ( {\bf r} ) = - \frac{\hbar^2}{2m} \nabla^2 + V ( {\bf r} ) - \mu.
\end{eqnarray}

\noindent
Here we employ the following Bogoliubov-Popov approximation:

\begin{eqnarray}
 \hat{\psi}^{\dagger} ( {\bf r} ) \hat{\psi} ( {\bf r} ) \hat{\psi} ( {\bf r} )
  & \simeq & 2 \langle \hat{\psi}^{\dagger} ( {\bf r} )
                        \hat{\psi} ( {\bf r} ) \rangle  \hat{\psi} ( {\bf r} )
                           \nonumber \\
  & & + \langle \hat{\psi} ( {\bf r} ) \hat{\psi} ( {\bf r} ) \rangle
                        \hat{\psi}^{\dagger} ( {\bf r} )\nonumber \\
 \hat{\psi}^{\dagger} ( {\bf r} ) \hat{\psi}^{\dagger} ( {\bf r} )
    \hat{\psi} ( {\bf r} ) \hat{\psi} ( {\bf r} )
  & \simeq & 4 \langle \hat{\psi}^{\dagger} ( {\bf r} )
                        \hat{\psi} ( {\bf r} ) \rangle
    \hat{\psi}^{\dagger} ( {\bf r} ) \hat{\psi} ( {\bf r} ) \nonumber \\
  & & + \langle \hat{\psi}^{\dagger} ( {\bf r} )
                 \hat{\psi}^{\dagger} ( {\bf r} ) \rangle
                 \hat{\psi} ( {\bf r} ) \hat{\psi} ( {\bf r} ) \nonumber \\
  & & + \langle \hat{\psi} ( {\bf r} ) \hat{\psi} ( {\bf r} ) \rangle
        \hat{\psi}^{\dagger} ( {\bf r} ) \hat{\psi}^{\dagger} ( {\bf r} )
\nonumber \\
 \hat{\psi}^{\dagger} ( {\bf r} ) \hat{\psi} ( {\bf r} ) \hat{\psi} ( {\bf r} )
    & \simeq & 2 \rho ( {\bf r} ) \hat{\psi} ( {\bf r} ) \nonumber\\
 \hat{\psi}^{\dagger} ( {\bf r} ) \hat{\psi}^{\dagger} ( {\bf r} )
    \hat{\psi} ( {\bf r} ) \hat{\psi} ( {\bf r} )
    & \simeq & 4 \rho ( {\bf r} )
        \hat{\psi}^{\dagger} ( {\bf r} ) \hat{\psi} ( {\bf r} ) \nonumber
\end{eqnarray}

\noindent
 with the non-condensate density defined as:
 $\langle \hat{\psi}^{\dagger} ( {\bf r} ) \hat{\psi} ( {\bf r} ) \rangle
    = \rho ( {\bf r} ).$
 Thus we arrive at the following effective one-body Hamiltonian within the
 present mean-field approximation:

\begin{eqnarray}
 \hat{H}_{rot} 
   & = & \int d {\bf r} \;
      [ \; \{ \; \phi^{\ast} ( {\bf r} ) h ( {\bf r} ) \phi ( {\bf r} )
       + \frac{1}{2} g { | \phi ( {\bf r} ) | }^4 \; \}   \nonumber \\
   & + & \hat{\psi}^{\dagger} ( {\bf r} )
       \{ \; h ( {\bf r} ) + g | \phi ( {\bf r} ) | ^2
          + 2g \rho ( {\bf r} ) \; \}
      \phi ( {\bf r} ) + h.c. \nonumber \\
   & + & \hat{\psi}^{\dagger} ( {\bf r} )
       \{ \; h ( {\bf r} ) + 2g | \phi ( {\bf r} ) | ^2
          + 2g \rho ( {\bf r} ) \; \}
      \hat{\psi} ( {\bf r} ) \nonumber \\
   & + & \frac{g}{2}
      \hat{\psi}^{\dagger} ( {\bf r} ) \hat{\psi}^{\dagger} ( {\bf r} )
      \phi ( {\bf r} ) \phi ( {\bf r} ) + h.c. \nonumber \\
   & + & i \hbar
      \{ \; \phi^{\ast} ( {\bf r} ) {\vec \omega} \cdot ( {\bf r} \times
         \nabla ) \phi ( {\bf r} ) \nonumber \\
   & & + \hat{\psi}^{\dagger} ( {\bf r} ) {\vec \omega} \cdot ( 
           {\bf r} \times 
         \nabla ) \phi ( {\bf r} ) \nonumber \\
   & & + \phi^{\ast} ( {\bf r} ) {\vec \omega} \cdot ( {\bf r} \times
         \nabla ) \hat{\psi} ( {\bf r} ) \nonumber \\
   & & + \hat{\psi}^{\dagger} ( {\bf r} ) {\vec \omega} \cdot ( 
         {\bf r} \times
         \nabla ) \hat{\psi} ( {\bf r} )
      \; \} \; ]
\end{eqnarray}

\noindent
This can be diagonalized by the usual Bogoliubov transformation by noting
the commutation relation:

\begin{eqnarray}
 [ \; \hat{\psi} ( {\bf r}^\prime ) , \hat{H}_{rot} \; ]
   = & \{ &  h ( {\bf r} ) + g | \phi ( {\bf r} ) | ^2
         + 2g \rho ( {\bf r} ) \nonumber \\
     & & +  i \hbar {\vec \omega} \cdot ( {\bf r} \times \nabla ) \; \}
                       \phi ( {\bf r} ) \nonumber \\
     & + &  h ( {\bf r} ) \hat{\psi} ( {\bf r} )
         + 2g | \phi ( {\bf r} ) | ^2 \hat{\psi} ( {\bf r} )  \nonumber \\
     & + & g \phi^2 ( {\bf r} ) \hat{\psi}^{\dagger} ( {\bf r} )
         + 2g \rho ( {\bf r} ) \hat{\psi} ( {\bf r} )  \nonumber \\
     & - & i \hbar {\vec \omega} \cdot
            [ \; \nabla \hat{\psi} ( {\bf r} )
                                \times {\bf r} \; ].
\end{eqnarray}

\noindent
The Bogoliubov transformation is given by

\begin{eqnarray}
 \hat{\psi} ( {\bf r} ) = \sum_{\bf q}
     [ \; u_{\bf q} ( {\bf r} ) \eta_{\bf q}
       -  v_{\bf q}^{\ast} ( {\bf r} ) \eta_{\bf q}^{\dagger} \; ].
\end{eqnarray}

\noindent
The resulting diagonalized form reads

\begin{eqnarray}
  \hat{H}_{rot} = E_0
     + \sum_{\bf q} \varepsilon_{\bf q} \eta_{\bf q}^{\dagger} \eta_{\bf q}.
\end{eqnarray}

\noindent
This diagonalization can be accomplished under the following conditions.
First we impose the condition on $\phi(\bf r)$ as

\begin{eqnarray}
 \{ \; h ( {\bf r} ) & + & g | \phi ( {\bf r} ) | ^2
         + 2g \rho ( {\bf r} )
         +  i \hbar {\vec \omega}{\cdot} ( {\bf r} \times \nabla ) 
         \; \} \phi ( {\bf r} ) = 0
\end{eqnarray}

\noindent
which is nothing but the so-called Gross-Pitaevskii (GP) equation
containing the non-condensate contribution.

\noindent
We must fulfill the following equations for diagonalization:

\begin{eqnarray}
 \{ \; h ( {\bf r} ) + 2g | \phi ( {\bf r} ) | ^2
  & + & 2g \rho ( {\bf r} ) \; \} u_{\bf q} ( {\bf r} )
      - g \phi^2 ( {\bf r} ) v_{\bf q} ( {\bf r} ) \nonumber \\
  & - & i \hbar {\vec \omega} \cdot \{ \nabla u_{\bf q} ( {\bf r} ) \times
         {\bf r} \} = \varepsilon_{\bf q} u_{\bf q} ( {\bf r} ) \\
 \{ \; h ( {\bf r} ) + 2g | \phi ( {\bf r} ) | ^2
  & + & 2g \rho ( {\bf r} ) \; \} v_{\bf q} ( {\bf r} )
      - g  \phi^{\ast 2} ( {\bf r} ) u_{\bf q} ( {\bf r} ) \nonumber \\
  & + & i \hbar {\vec \omega} \cdot \{ \nabla v_{\bf q} ( {\bf r} )
         \times {\bf r} \} = - \varepsilon_{\bf q} v_{\bf q} ( {\bf r} ).
\end{eqnarray}

\noindent
This coupled equation is known as Bogoliubov-de Gennes (BdG)
equation\cite{de gennes}
 used both for BEC and for superconductors.

 We notice that the eigenfunctions; $u_{\bf q} ( {\bf r} ) $ and $v_{\bf q}
( {\bf r} ) $
 satisfy the otho-normalization requirement
 or the positive norm condition:

\begin{eqnarray}
 \int d {\bf r} [ \; u_{\bf p}^{\ast} ( {\bf r} ) u_{\bf q} ( {\bf r} )
      - v_{\bf p}^{\ast} ( {\bf r} ) v_{\bf q} ( {\bf r} ) \; ]
      = \delta_{{\bf p},{\bf q}}.
\end{eqnarray}

\noindent
The non-condensate density is determined self-consistently by
\begin{eqnarray}
 \rho ( {\bf r} ) & = &
      \langle \hat{\psi}^{\dagger} ( {\bf r} )
             \hat{\psi} ( {\bf r} ) \rangle  \nonumber \\
  & = & \sum_{{\bf q}}[ \; | u_{\bf q} ( {\bf r} ) | ^2 f ( \epsilon_{\bf q} )
      + | v_{\bf q} ( {\bf r} ) | ^2 \{ f ( \epsilon_{\bf q} ) + 1 \} \; ].
\end{eqnarray}
These equations (9), (10) and (11) constitute a complete set of the
non-linear coupled equations.

\subsection{Cylindrical system}

In order to describe a single vortex threading through the center
of the cylindrical symmetric system, it is convenient to use a
cylindrical coordinate: ${\bf r} = ( r , \theta , z )$.
We impose the boundary condition that  the condensate wave function
and the eigenfunctions for the non-condensate vanish at  $r=R$, namely,
$ \phi ( r = R ) = 0 , u_{\bf q} ( r = R ) = 0 $, and $v_{\bf q} ( r = R ) =
0 $,
where  $R$ is taken far enough from the center.

\noindent
The confining potential is given by
\begin{eqnarray}
 V ( {\bf r} ) = \frac{1}{2} m ( 2 \pi \nu_r )^2 r^2
\end{eqnarray}
and the periodic boundary condition is employed along the $z$ direction.
The condensate wave function is written in the form:
\begin{eqnarray}
 \phi ( r,\theta,z ) & = & \phi ( r ) e^{i w \theta}.
\end{eqnarray}
The winding number $w$ is finite when a vortex is present and vanishes
when a vortex is absent. In the following we consider the vortex with $w=1$
only.

The eigenfunctions are also written by the symmetry reasons as
\begin{eqnarray}
 u_{\bf q} ( {\bf r} ) & = & u_{\bf q} ( r )
            e^{i q_z z} e^{i ( q_{\theta} + w ) \theta} \\
 v_{\bf q} ( {\bf r} ) & = & v_{\bf q} ( r )
            e^{i q_z z} e^{i ( q_{\theta} - w ) \theta}.
\end{eqnarray}

\noindent
The eigenvalues
 ${\bf q}= ( q_r , q_{\theta} , q_z ) $ may take the following values:
 $ q_r =1,2,3,, \cdots$,
 $q_{\theta}=0 , \pm 1 , \pm 2 , , \cdots$, and
  $  q_z=0, \pm 2 \pi / L , \pm 4 \pi / L , \cdots$,
where $L$ is the period of the length along the $z$-axis.

By choosing the external rotation ${\vec \omega} = ( 0,0,\omega)$,
 we obtain the GP equation as

\begin{eqnarray}
 [ & - & \frac{\hbar^2}{2 m}
       \{ \frac{d^2}{d r^2} + \frac{1}{r} \frac{d}{dr}
          - \frac{w^2}{r^2} \; \} - \mu + V ( r ) \nonumber \\
   & + & g \phi^2 ( r ) + 2 g \rho ( r )
    - \hbar \omega w \; ] \phi ( r ) = 0.
\end{eqnarray}
Similarly, the BdG equations are now

\begin{eqnarray}
 & [ & \; - \frac{\hbar^2}{2 m} \;
    \{ \; \frac{d^2}{d r^2} + \frac{1}{r} \frac{d}{dr}
          - \frac{(q_{\theta} + w )^2}{r^2} - q_{z}^2 \; \}
          - \mu + V ( r ) \nonumber \\
 & &  + 2 g ( \phi^2 + \rho )
          - \hbar \omega ( w + q_{\theta} ) \; ] u_{\bf q} ( r )
               \nonumber \\
 & & - g \phi^2 ( r ) v_{\bf q} ( r )
          = \varepsilon_{\bf q} u_{\bf q} ( r ) \\
 & [ & \; - \frac{\hbar^2}{2 m} \;
    \{ \; \frac{d^2}{d r^2} + \frac{1}{r} \frac{d}{dr}
          - \frac{(q_{\theta} - w )^2}{r^2} - q_{z}^2 \; \}
          - \mu + V ( r ) \nonumber \\
 & &  + 2 g ( \phi^2 + \rho )
          + \hbar \omega ( w - q_{\theta} ) \; ] v_{\bf q} ( r )
            \nonumber \\
 & & - g \phi^2 ( r ) u_{\bf q} ( r ) 
          = \varepsilon_{\bf q} v_{\bf q} ( r ).
\end{eqnarray}

\noindent
The actual implimentation of the numerical computation for these
coupled equations is described before\cite{isoshima3,isoshima4}.

\subsection{Free energy}

The internal energy $E$ of the system is calculated by taking the
Hamiltonian
within the mean-field approximation, namely,

\begin{eqnarray}
 E & = & \langle \hat{H}_{rot} \rangle + \mu N \nonumber \\
   & = & \int d {\bf r} \;
  [ \;
   \{ \; \phi^{\ast} ( {\bf r} ) h ( {\bf r} ) \phi ( {\bf r} )
       + \frac{1}{2} g { | \phi ( {\bf r} ) | }^4 \; \}
           \nonumber \\
   & & + \langle \hat{\psi}^{\dagger} ( {\bf r} )
       \{ \; h ( {\bf r} ) + g | \phi ( {\bf r} ) | ^2
          + 2g \rho ( {\bf r} ) \; \}
      \phi ( {\bf r} ) \rangle + h.c. \nonumber \\
   & & + \langle \hat{\psi}^{\dagger} ( {\bf r} )
       \{ \; h ( {\bf r} ) + 2g | \phi ( {\bf r} ) | ^2
          + 2g \rho ( {\bf r} ) \; \}
      \hat{\psi} ( {\bf r} ) \rangle \nonumber \\
   & & + \frac{g}{2}
      \langle \hat{\psi}^{\dagger} ( {\bf r} ) \hat{\psi}^{\dagger} ( {\bf r} )
      \phi ( {\bf r} ) \phi ( {\bf r} ) \rangle + h.c. \nonumber \\
   & & + i \hbar
      \{ \; \langle \phi^{\ast} ( {\bf r} ) {\vec \omega} \cdot ( 
         {\bf r} \times
         \nabla ) \phi ( {\bf r} ) \rangle
         + \langle \hat{\psi}^{\dagger} ( {\bf r} )
         {\vec \omega} \cdot ( {\bf r} \times
         \nabla ) \phi ( {\bf r} ) \rangle \nonumber \\
        & &
         + \langle \phi^{\ast} ( {\bf r} ) {\vec \omega} \cdot ( 
         {\bf r} \times
         \nabla ) \hat{\psi} ( {\bf r} ) \rangle
         + \langle \hat{\psi}^{\dagger} ( {\bf r} )
         {\vec \omega} \cdot ( {\bf r} \times
         \nabla ) \hat{\psi} ( {\bf r} ) \rangle
      \; \} \; ] \nonumber \\ 
    & & + \mu N
\end{eqnarray}
or, it can be written as follows:

\begin{eqnarray}
 E & = & \int d {\bf r} \;
  [ \;
   \{ \; \phi^{\ast} ( {\bf r} ) h ( {\bf r} ) \phi ( {\bf r} )
       + \frac{1}{2} g { | \phi ( {\bf r} ) | }^4 \; \}
          \nonumber \\
   & & +  \langle \hat{\psi}^{\dagger} ( {\bf r} )
       \{ \; h ( {\bf r} ) + 2g | \phi ( {\bf r} ) | ^2
          + 2g \rho ( {\bf r} ) \; \}
      \hat{\psi} ( {\bf r} )\rangle \nonumber \\
   & & + \frac{g}{2}
      \langle \hat{\psi}^{\dagger} ( {\bf r} ) \hat{\psi}^{\dagger} ( {\bf r} )
      \phi ( {\bf r} ) \phi ( {\bf r} ) \rangle + h.c. \nonumber \\
   & & + i \hbar
      \{ \phi^{\ast} ( {\bf r} ) {\vec \omega} \cdot ( {\bf r} \times
         \nabla ) \phi ( {\bf r} ) \nonumber \\
   & & + \langle \hat{\psi}^{\dagger} ( {\bf r} )
         {\vec \omega} \cdot ( {\bf r} \times
         \nabla ) \hat{\psi} ( {\bf r} ) \rangle
      \} \; ] + \mu N \nonumber \\
   & = & E_0 + \mu N_0 + \langle \hat{H}_{1} \rangle + \mu N_{nc}
\end{eqnarray}

\noindent
where we define
\begin{eqnarray}
 E_0
   & = & \int d {\bf r} \;
    [ \;
       \phi^{\ast} ( {\bf r} ) h ( {\bf r} ) \phi ( {\bf r} )
       + \frac{1}{2} g { | \phi ( {\bf r} ) | }^4 \nonumber \\
   & & + i \hbar \phi^{\ast} ( {\bf r} ) {\vec \omega} \cdot ( {\bf r} \times
         \nabla ) \phi ( {\bf r} ) \; ] \\
 \hat{H}_{1}
   & = & \int d {\bf r} \;
    [ \;
       \hat{\psi}^{\dagger} ( {\bf r} )
       \{ \; h ( {\bf r} ) + 2g | \phi ( {\bf r} ) | ^2
       + 2g \rho ( {\bf r} ) \; \} \hat{\psi} ( {\bf r} ) \nonumber \\
   & & + \frac{g}{2}
       \hat{\psi}^{\dagger} ( {\bf r} ) \hat{\psi}^{\dagger} ( {\bf r} )
       \phi ( {\bf r} ) \phi ( {\bf r} ) 
       + \frac{g}{2}
       \hat{\psi} ( {\bf r} ) \hat{\psi} ( {\bf r} )
       \phi^{\ast} ( {\bf r} ) \phi^{\ast} ( {\bf r} ) \nonumber \\
   & & + i \hbar \hat{\psi}^{\dagger} ( {\bf r} )
         {\vec \omega} \cdot ( {\bf r} \times
         \nabla ) \hat{\psi} ( {\bf r} ) \; ].
\end{eqnarray}
In terms of the cylindrical coordinate ${\bf r} = ( r , \theta , z )$, we
obtain

\begin{eqnarray}
 E_0
   & = & \int d {\bf r} \;
    [ \;
       \phi^{\ast} ( r ) \;
       \{ \; - \frac{\hbar^2}{2m} \nabla^2 + V ( r ) - \mu
         \; \} \; \phi ( r ) \nonumber \\
   & & + \frac{1}{2} g { | \phi ( r ) | }^4
       + - \hbar \omega w \phi^2 ( r ) \; ] \nonumber \\
   & = & \int d {\bf r} \;
    [ \; \phi^{\ast} ( r ) \; ( - \frac{\hbar^2}{2m} )
       \{ \; \frac{d^2}{d r^2}
       + \frac{1}{r} \frac{d}{d r} - \frac{w^2}{r^2}
       \; \} \phi ( r ) \nonumber \\ 
   & & + V ( r ) \phi^2 ( r ) 
       - \mu \phi^2 ( r ) + \frac{1}{2} g \phi^4 ( r )
       - \hbar \omega w \phi^2 ( r ) \; ].
\end{eqnarray}

\noindent
As shown in Appendix,
the total internal energy is now evaluated by the formula:
\begin{eqnarray}
 E & = & \langle \hat{H}_{rot} \rangle + \mu N \nonumber \\
   & = & 2 \pi L \int d r
    [ \; \phi^{\ast} ( r ) \; ( - \frac{\hbar^2}{2m} )
       \{ \; \frac{d^2}{d r^2}
       + \frac{1}{r} \frac{d}{d r} - \frac{w^2}{r^2}
       \; \} \phi ( r ) \nonumber \\
   & & + V ( r ) \phi^2 ( r ) 
       + \frac{1}{2} g \phi^4 ( r )
       - \hbar \omega w \phi^2 ( r ) \; ] \nonumber \\
   & & + \sum_{\bf q} \varepsilon_{\bf q} f ( \varepsilon_{\bf q} )
       + \sum_{\bf q} \varepsilon_{\bf q} \int d {\bf r} | v_{\bf q} |^2
       + \mu \int d {\bf r} \rho ( {\bf r} )
\end{eqnarray}

\noindent
where $f( \varepsilon ) = 1/\{ \exp ( \varepsilon_{\bf q} / k_B T ) - 1\}$ 
is the Bose distribution function.

The entropy $S$ of the system is obtained 
by the following expression\cite{Giorgini}:
\begin{eqnarray}
 S = k_B \sum_{\bf q} 
      \{ \;
        \beta \varepsilon_{\bf q} f ( \varepsilon_{\bf q} )
        - \ln [ \; 1 - \exp ( - \varepsilon_{\bf q} / k_B T ) \; ]
      \; \}. 
\end{eqnarray}

\noindent
From eqs. (26) and (27), we can estimate the free energy $F=E-TS$
of the vortex and non-vortex states.

\subsection{Calculated system}

In order to perform the self-consistent calculation to determine
various characteristic frequencies as a function $T$, we must  specify the
parameters: As a  typical example, we take up the vortex nucleation
experiment on
$^{87}Rb$ atom vapor by Chevy et al\cite{chevy}.
The scattering length $a=5.5\times10^{-9}m$ and atom mass
$m=1.44\times10^{-25}kg$
for $^{87}$Rb atom. The radial and $z$ axis trapping frequencies are 175Hz
and 10.3Hz
respectively. We evaluate the density profile of the condensate in Fig.1 by
solving the GP equation at $T$=0 for the total number $N=2.5\times 10^{5}$.
The area density along the $z$ axis can be
estimated as $n_z=2\times10^9/m\sim3\times10^9/m$ at the center of $r=0$
as seen from Fig.1(b). In the following the dimensionless ``gas parameter''
$an_z=11$
is  chosen in the present two-dimensional disk system and the rotation
frequency
is normalized by the radial confining potential
$\omega_{\perp}\equiv2\pi\nu_r$,
denoted  by $\Omega\equiv{\omega\over \omega_{\perp}}$ from now on
($\omega_{\perp}/2\pi=175Hz$).
The system size is set to $R=7\mu m$ and $L=5\mu m$.
The transition temperature of the present stsem $T_{c0}=207.8nK$ at $\Omega=0$.
The energy is scaled by $\hbar\omega_{\perp}=h\nu_r$.

\section{Local stability}
\subsection{Local stability of the vortex state}
The local stability of the single vortex state is defined by the  frequency
region bound by the two frequencies: $\Omega^L_{local}<\Omega<\Omega^U_{local}$
where the excitation spectrum is well-defined and no negative frequency exists.
We show in Fig.2 the density profiles for the condensate (a) and
non-condensate (b)
and also the associated excitation spectrum (c) for various $\Omega$
values under a fixed $T$ ($T/T_{c0}=0.4$).
As $\Omega$ increases, we can see from Fig.2(a) that the condensate is
depressed from the outside region while the non-condensate increases its
density
at the vortex center, filling the vortex core by the non-condensate,
eventually a large
amount of the non-condensate becomes piling up the outside region for larger
$\Omega$ values.
The filling-in of the non-condensate effectively helps pushing
up the so-called
anomalous mode, that is, the
negative eigenvalue at $q_{\theta}=-1$ toward a positive one as seem from
Fig.2(c),
resulting in the local stability of the vortex state and giving rise to
$\Omega^L_{local}$\cite{salomaa}.
In Fig. 2(a) as the inset we display  the trace of the negative mode and
also the lowest
mode at $q_{\theta}>0$ which becomes negative at $\Omega^U_{local}$
at higher $\Omega$'s. It is seen that for
$\Omega^L_{local}<\Omega<\Omega^U_{local}$
the excitation spectrum is well defined, implyig that a vortex is locally
stable in the
energy configuration space.

The piling-up of the non-condensate at the outside region arises from the
repulsive interaction between the condensate and non-condensate parts.
It is rather remarkable to notice from Fig.2(a) in this connection that the
vortex core size measured by the peak position of the condensate density
is hardly changed under varying $\Omega$.

As $T$ is varied under a fixed  $\Omega$, both density profiles of the
condensate
and non-condensate change into each other shown in Figs.3 (a) and (b)
respectively.
As seen from Fig.3(a) the core radius defined by the maximum position of the
density profile of the condensate slightly increases as $T$ increases.
The non-condensate density quickly fills in the central core region and also
the outside region
as $T$ increases as seen from Fig.3(b).
Again the the negative eigenmode at $q_{\theta}=-1$ becomes positive at
$\Omega^L_{local}$, shown in Fig. 3(b) as the inset. It will be seen shortly
that $\Omega^L_{local}$
is a monotonically decreasing function of $T$.

When we perform the self-consistent calculation,
we exclude modes with negative energy in eq.(13).
The negative mode(s) represent the instability of system consist of
condensate and non-condensate.
The steep step in the insets of Fig.2(a) and Fig.3(b) appears
when the lowest mode becomes ignored.

\subsection{Local stability of the non-vortex state}

The corresponding local stability can be examined for the non-vortex state
in a similar way: We show the corresponding example for that case
in Fig.4 where (a) and (b) exhibit the density
profiles for the condensate and non-condensate respectively.
Again as $\Omega$ increases, the former (latter) is depleted (piled up)
from the outside region. There is no filling-in effect of the non-condensate
in this
vortex-free state.

The excitation spectrum as depicted in Fig.4(c) shows
the appearance of the negative mode at  $\Omega_{w=0}$
for certain mode
with $q_{\theta}>0$. This signals an intrinsic instability of this
non-vortex state
toward some other state, most likely toward the vortex state in the present
situation.
The trace of the lowest mode with $q_{\theta}>0$ is shown in the 
inset of Fig.4(a) which
is seen to
$\Omega_{w=0}  =0.566$ in this case.

The $T$-dependent poperties of this vortex-free state are depicted in Fig.
5: As $T$ increases,
the density
conversion from the condensate to the non-condensate occurs. This is shown
in Figs. 5(a) and (b)
where each density distribution is displayed.
The inset of Fig.5(a) exhibits the trace of the lowest mode with
$q_{\theta}>0$,
indicating that the negative mode in $q_{\theta}>0$ become positive
at $T/T_{c0}=0.4$ in this example.

\section{Global stability}

The relative energies of the non-vortex and the vortex states should change
as functions
of $\Omega$ and $T$. Lundh et al\cite{lundh} estimate the energy increment
$\varepsilon$
per unit length due to the vortex creation for the uniform system at $T=0$ as

\begin{eqnarray}
\varepsilon=\pi n{\hbar^2\over m}\ln{1.464b\over \xi}
\end{eqnarray}

\noindent
where $n$ is the condensate density, $b$ is the radius of a bucket and the
coherent length
$\xi=1/8\pi an$. In the presence of the confining potential
the above expression is modified  to

\begin{eqnarray}
\varepsilon=\pi n_0{\hbar^2\over m}\ln{0.888R\over \xi_0}
\end{eqnarray}

\noindent
with $n_0$ the peak density, $R$ the trapping radius and $\xi_0=1/8\pi an_0$.
The angular momentum $L$ per unit length is given by
$L=\hbar N$, thus the critical frequency $\Omega^L_{global}$
is determined by

\begin{eqnarray}
\varepsilon-\Omega^L_{global}L=0.
\end{eqnarray}

\noindent
Under the Thomas-Fermi approximation this leads to

\begin{eqnarray}
\Omega^L_{global}=2{\hbar\over mR^2}\ln{0.888R\over \xi_0}
\end{eqnarray}

\noindent
which indicates the global stability of a vortex relative to the non-vortex
state,
where $R$ is the radius determined by the Tomas-Fermi
approximation.
At $T$=0 we can easily estimate $\Omega^L_{global}=0.372$
for the Thomas-Fermi approximation, which is in good agreement with our own
calculation $\Omega^L_{global}=0.382$.
Stringari\cite{stringari2} further extends it to finite temperatures by
neglecting the
non-condensate contribution. We have performed the fully
self-consistent calculation, taking into account the non-condensate
contribution.

We compare various approximate estimates $\Omega^L_{global}$
as a function of  $T$. First the Stringari's $\Omega^L_{global}$ is
well reproduced by our numerical calculation using the expression (31)
where $n_0(T)$
and $R(T)$  are calculated by our self-consistent solution.
Starting with eq.(14)
and equating the corresponding non-vortex energy, we can evaluate
$\Omega^L_{global}$
as shown in Fig.6 where curve (2) indicates the result
when neglecting the non-condensate contribution while (3) is fully
consistent results.
It is seen that as $T$ increases two curves deviate progressively showing
the importance of
the non-condensate contribution.
However, curve (3) neglects the effects of the entropy S defined by eq.(27).
In fact, the global stability of the vortex state relative to 
the non-vortex state at finit temperature is defined by 
the comparison of the two free energies in the system 
with and without a vortex.
The $T$-dependence of the critical value is plotted as curve (4) in Fig.6.

\section{Phase diagram and remarks}

We show in Fig.7 the four characteristic frequencies;
$\Omega^L_{local}$,
$\Omega^L_{global}$, $\Omega_{w=0}$ and $\Omega^U_{local}$
together with the critical temperature $T_c(\Omega)$.  This is estimated
as

\begin{eqnarray}
T_c(\Omega)=(1-\Omega^2)^{1\over 3}
\end{eqnarray}

\noindent
by Stringari\cite{stringari2} for a non-interacting system, which is also
approximately valid
for the present dilute Bose gas. It is seen that $\Omega^L_{local}$ is
a decreasing function of $T$
because the non-condensate fraction excited thermally serves as stabilizing a
vortex,
a fact pointed out already by Isoshima and Machida\cite{isoshima2}, and also
recently
by Virtanen et al\cite{salomaa}.
$\Omega^L_{global}(T)$ is almost $T$-independent up to 0.5$T_{c0}$ and then
quickly increases toward the $T_c(\Omega)$ line. According to eq.(31) they
meet tangentially.

As we have discussed on III,
the instability of the non-vortex and the single vortex state
is defined by the same reason.
However, as seen from Fig.7,
the critical value of the instability the critical $\Omega_{w=0}$,
which correspond to the instability of the non-vortex state,
disagrees with the critical $\Omega_{local} ^U$.
We interpret the disagreement by the following effect.
From the eigenvalue equations (19) and (20), we define 
the effective potential
\begin{eqnarray}
 V_{eff} ( r ) \; \equiv \; V ( r ) + 2g \; [ \; 
          | \phi ( r ) |^2 + \rho ( r ) \; ].
\end{eqnarray}
In comparison with the non-vortex state,
the density distribution of the condensate $| \phi ( r ) |^2$
in the vortex state is slightly pushed out because of 
the angular momentum.
As seen from Fig.8(a), this situation is fully understood in terms of
$V_{eff} ( r )$.
As discussed in III, the eigenvalues with $q_{\theta}>0$ 
move down as $\Omega$ increases.
This means that the non-condensate quickly grows at the edge of 
the condensate, and the non-vortex state (the single vortex state)
becomes unstable at $\Omega_{w=0}$ ($\Omega_{local} ^U$).
In short, the higher $\Omega$ requires in order for the non-condensate
to increase at the edge of the condensate in the vortex state,
and we find that $\Omega_{local} ^U$ is always higher than
$\Omega_{w=0}$.  

The remaining $\Omega_{w=0}$ and $\Omega^U_{local}$ are a rather
strong increasing function of $T$. As mentioned before, $\Omega_{w=0}$ turns
out to
be very near the observed nucleation frequency by Chevy et al\cite{chevy},
which is independent of the details of the experimental parameters used.
They find $\Omega_{nuc}\sim0.68\omega_{\perp}$. This is compared with our
$\Omega_{w=0}=0.55$, which is nearer than the thermodynamic critical frequency
$\Omega^L_{global}\sim 0.39$ at $T=0$.
This may be pure accident, but we can check it by quantitatively examining
it through
its $T$ dependence.

It is important to notice that $\Omega^U_{local}(T)$ is not much
different from  $\Omega_{w=0}(T)$ in its value as seen from Fig.7.
Thus once a vortex nucleates
at $\Omega_{w=0}(T)$, slight increase of $\Omega$ brings multiple vortices at
$\Omega^U_{local}(T)$, that is, the frequency window of the single vortex
state is rather narrow.
This is  indeed an experimental fact
\cite{chevy}($\Omega_{nuc}/\omega_{\perp}=0.668$,
$\Omega_{multi-vortices}/\omega_{\perp}=0.697$).

If the system is kept rotating during cooling from above $T_c$, then the
nucleation
likely begins when $\Omega$ reaches $\Omega^L_{local}$. If the system, cooled
already, on the
other hand, begins rotating, then at $\Omega_{w=0}$ a vortex may appear.
It is known for some time that in superfluid $^4$He the former is lower than
the latter as expected while in BEC both nucleation protocols give rise to
almost same nucleation frequency.
This is quite understandable because two frequencies; $\Omega^L_{local}$  and
$\Omega_{w=0}$ become almost equal at $T=0$ when the interaction is weak while
these two frequencies become quite distinctive when the interaction is strong
(see Fig.3 in Isoshima and Machida\cite{isoshima4}).
It is also true for finite temperatures that the two characteristic frequencies
move oppositely upon increasing $T$ as seen from Fig.7.
Here there is a good chance to directly confirm the characteristic nucleation
frequencies for two different cooling protocols, a situation quite
different from the superfluid
$^4$He rotating bucket case where the rough surface of a bucket prevents
us from knowing the intrinsic nucleation frequency. The present gaseous system
is free from this problem.

Quite recentry,
yet another vortex experiment is reported by Abo-Shaeer 
et al\cite{abo}
where many vortices over 100 are observed.
The very low nucleation frequency $\Omega_{nuc} \simeq 0.25\omega_{\perp}$
in their experiment is qualitatively understandable because 
the gas parameter $an_z$ is large (an order of 500),
implying that the characteristic frequencies $\Omega_{w=0}$,
$\Omega_{global}^L$ and $\Omega_{local}^L$ are all small
in that region (see again Fig.3 in Isoshima and Machida \cite{isoshima4}).

\section*{Acknowledgments}
The authors thank K.W. Madison for detailed discussions on
their vortex nucleation experiments and also W. F. Vinen for
telling us the situation of the vortex nucleation problem in superfluid
$^4$He.

\appendix

\section*{}

We first express each term of eq. (4) in terms of
$\eta^{\dagger}$ and $\eta$ as

\begin{eqnarray}
 \hat{\psi}^{\dagger} ( {\bf r} ) \{ \; h ( {\bf r} )
  & + & 2g | \phi ( {\bf r} ) | ^2
        + 2g \rho ( {\bf r} ) \; \} \hat{\psi} ( {\bf r} ) \nonumber \\
  & = & \sum_{i,j} [ \;
        \eta_{i}^{\dagger} \eta_{j} \;
        \{ \; u_{i}^{\ast} ( h + 2g | \phi | ^2 + 2g \rho ) u_{j} \; \}
\nonumber \\
  & &   + \eta_{i} \eta_{j}^{\dagger} \;
        \{ \; v_{i} ( h + 2g | \phi | ^2 + 2g \rho ) v_{j}^{\ast} \; \}
\nonumber \\
  & &   - \eta_{i}^{\dagger} \eta_{j}^{\dagger} \;
        \{ \; u_{i}^{\ast}
          ( h + 2g | \phi | ^2 + 2g \rho ) v_{j}^{\ast} \; \} \nonumber \\
  & &   - \eta_{i} \eta_{j} \;
        \{ \; u_{i}^{\ast}
          ( h + 2g | \phi | ^2 + 2g \rho ) u_{j} \; \} \; ] \nonumber \\
 \hat{\psi}^{\dagger} ( {\bf r} ) \hat{\psi}^{\dagger} ( {\bf r} )
  & = & \sum_{i,j} [ \;
        \eta_{i}^{\dagger} \eta_{j}^{\dagger} u_{i}^{\ast} u_{j}^{\ast}
        + \eta_{i} \eta_{j} v_{i} v_{j}
        - 2 \eta_{i}^{\dagger} \eta_{j} u_{i}^{\ast} v_{j} \; ] \nonumber \\
 \hat{\psi} ( {\bf r} ) \hat{\psi} ( {\bf r} )
  & = & \sum_{i,j} [ \;
        \eta_{i} \eta_{j} u_{i} u_{j}
        + \eta_{i}^{\dagger} \eta_{j}^{\dagger} v_{i}^{\ast} v_{j}^{\ast}
        - 2 \eta_{i} \eta_{j}^{\dagger} u_{i} v_{j}^{\ast} \; ] \nonumber
\end{eqnarray}
\begin{eqnarray}
 i \hbar 
  & \{ & \hat{\psi}^{\dagger} ( {\bf r} )
         {\vec \omega} \cdot ( {\bf r} \times \nabla )
         \hat{\psi} ( {\bf r} ) \; \} \nonumber \\
  & = & i \hbar \sum_{i,j} ( u_{i}^{\ast} \eta_{i}^{\dagger}
        - v_{i} \eta_{i} ) {\vec \omega} \cdot
        ( {\bf r} \times \nabla ) ( u_{j} \eta_{j}
        - v_{j}^{\ast} \eta_{j}^{\dagger} )  \nonumber \\
  & = & - i \hbar \sum_{i,j} ( u_{i}^{\ast} \eta_{i}^{\dagger}
        - v_{i} \eta_{i} ) {\vec \omega} \cdot
        [ \; \nabla ( u_{j} \eta_{j} - v_{j}^{\ast} \eta_{j}^{\dagger} )
        \times {\bf r} \; ] \nonumber \\
  & = & - i \hbar \sum_{i,j} {\vec \omega} \cdot
        [ \;
        \eta_{i}^{\dagger} \eta_{j} u_{i}^{\ast} ( \nabla u_{j} \times {\bf
r} )
        - \eta_{i}^{\dagger} \eta_{j}^{\dagger}
               u_{i}^{\ast} ( \nabla v_{j}^{\ast} \times {\bf r} ) \nonumber \\
  & &
        - \eta_{i} \eta_{j} v_{i} ( \nabla u_{j} \times {\bf r} )
        + \eta_{i} \eta_{j}^{\dagger} v_{i}
               ( \nabla v_{j}^{\ast} \times {\bf r} ). \nonumber
\end{eqnarray}

\noindent
We substitute these terms into eq.(4) and obtain
\begin{eqnarray}
 \hat{H}_{1}
  & = & \sum_{i,j} \int d {\bf r} \;
    [ \;
        \eta_{i}^{\dagger} \eta_{j} \; \{ \;
        u_{i}^{\ast} ( h + 2g | \phi | ^2 + 2g \rho ) u_{j} \nonumber \\
  & &   - \frac{g}{2} u_{i}^{\ast} v_{j} \phi^2
        - \frac{g}{2} u_{j} v_{i}^{\ast} \phi^{\ast 2}
        - i \hbar u_{i}^{\ast}
        {\vec \omega} \cdot ( \nabla u_{j} \times {\bf r} ) \; \} \nonumber \\
  & &   + \eta_{i} \eta_{j}^{\dagger} \; \{ \;
        v_{i}^{\ast} ( h + 2g | \phi | ^2 + 2g \rho ) v_{j} \nonumber \\
  & &   - \frac{g}{2} u_{j}^{\ast} v_{i} \phi^2
        - \frac{g}{2} u_{i} v_{j}^{\ast} \phi^{\ast 2}
        - i \hbar v_{i}^{\ast}
        {\vec \omega} \cdot ( \nabla v_{j} \times {\bf r} ) \; \} \nonumber \\
  & &   + ( \eta_{i} \eta_{j} term)
        + ( \eta_{i}^{\dagger} \eta_{j}^{\dagger}term) \; ]
\end{eqnarray}
where the coefficient of the $\eta_{i}^{\dagger} \eta_j$

\begin{eqnarray}
 u_{i}^{\ast} ( h + 2g | \phi | ^2 + 2g \rho ) u_{j}
  & &   - \frac{g}{2} u_{i}^{\ast} v_{j} \phi^2
        - \frac{g}{2} u_{j} v_{i}^{\ast} \phi^{\ast 2} \nonumber \\
  & &   - i \hbar u_{i}^{\ast}
        {\vec \omega} \cdot ( \nabla u_{j} \times {\bf r} )
\end{eqnarray}

\noindent
is rewritten by using

\begin{eqnarray}
\{ \; h + 2g | \phi |^2 
  & + & 2g \rho  \; \} u_{j}
        - g | \phi |^2 v_{j} \nonumber \\
  & &   - i \hbar {\vec \omega} \cdot ( \nabla u_{j} \times
         {\bf r} )  = \varepsilon_{j} u_{j}
\end{eqnarray}
\noindent
as
\begin{eqnarray}
 (8) & = & \varepsilon_{j} u_{i}^{\ast} u_{j}
           + g \phi^2 u_{i}^{\ast} v_{j}
           - \frac{g}{2} u_{i}^{\ast} v_{j} \phi^2
           - \frac{g}{2} u_{j} v_{i}^{\ast} \phi^{\ast 2} \nonumber \\
     & = & \varepsilon_{j} u_{i}^{\ast} u_{j}.
\end{eqnarray}

\noindent
Here we have utilized a relation:
\begin{eqnarray}
 \sum_{i,j} ( 
  & - & \frac{g}{2} u_{i}^{\ast} v_{j} \phi^2
           - \frac{g}{2} u_{j} v_{i}^{\ast} \phi^{\ast 2} ) \nonumber \\
  & &   = \sum_{i,j} \{ \; - \frac{g}{2} u_{i} (r ) v_{j} ( r ) \phi^2 ( r )
           - \frac{g}{2} u_{j} ( r )
           v_{i}^{\ast} ( r )  \phi^{\ast 2} ( r )  \; \} \nonumber \\
  & &   = \sum_{i,j} ( - g \phi^2 u_{i}^{\ast} v_{j} ).
\end{eqnarray}

\noindent
Similarly, the coefficient of the term $\eta\eta^{\dagger}$ in eq.(a7) is
found to be
\begin{eqnarray}
 - \sum_{i,j} \varepsilon_{j} v_{i}^{\ast} v_{j}. \nonumber
\end{eqnarray}

\noindent
Thus the expectation value of $\hat{H}_{1}$ is finally
evaluated as follows:

\noindent
\begin{eqnarray}
 \langle \hat{H}_{1} \rangle
  & = & \sum_{i,j} \int d {\bf r} [ \;
        \varepsilon_{j} u_{i}^{\ast} u_{j}
        \langle \eta_{i}^{\dagger} \eta_{j} \rangle
        - \varepsilon_{j} v_{i}^{\ast} v_{j}
        \langle \eta_{i} \eta_{j}^{\dagger} \rangle \nonumber \\
  & = & \sum_{i} \varepsilon_{i} \int d {\bf r} [ \;
        | u_{i} |^2 - | v_{i} |^2 \; ]
        \langle \eta_{i}^{\dagger} \eta_{j} \rangle
        - \sum_{i} \varepsilon_{i} \int d {\bf r} | v_{i} |^2 \nonumber \\
  & = & \sum_{i} \varepsilon_{i} \langle \eta_{i}^{\dagger} \eta_{j} \rangle
        - \sum_{i} \varepsilon_{i} \int d {\bf r} | v_{i} |^2
\end{eqnarray}

\noindent
with
\begin{eqnarray}
 \mu N_{nc}
  & = & \mu \int d {\bf r}
        \langle
        \hat{\psi}^{\dagger} ( {\bf r} ) \hat{\psi} ( {\bf r} )
        \rangle \nonumber \\
  & = & \mu \int d {\bf r} \rho ( {\bf r} ).
\end{eqnarray}


%

\begin{center}
 \leavevmode
 \epsfxsize=70mm
 \epsfbox{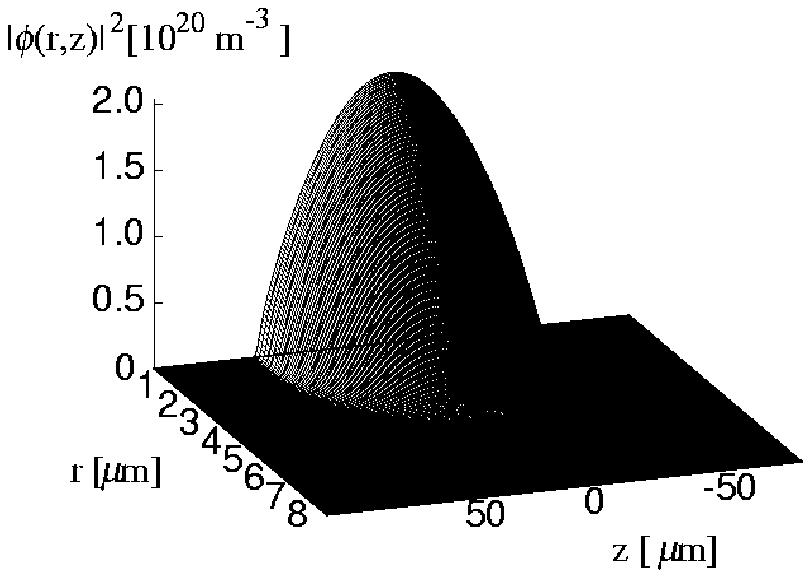}
\end{center}

(a)

\begin{center}
 \leavevmode
 \epsfxsize=70mm
 \epsfbox{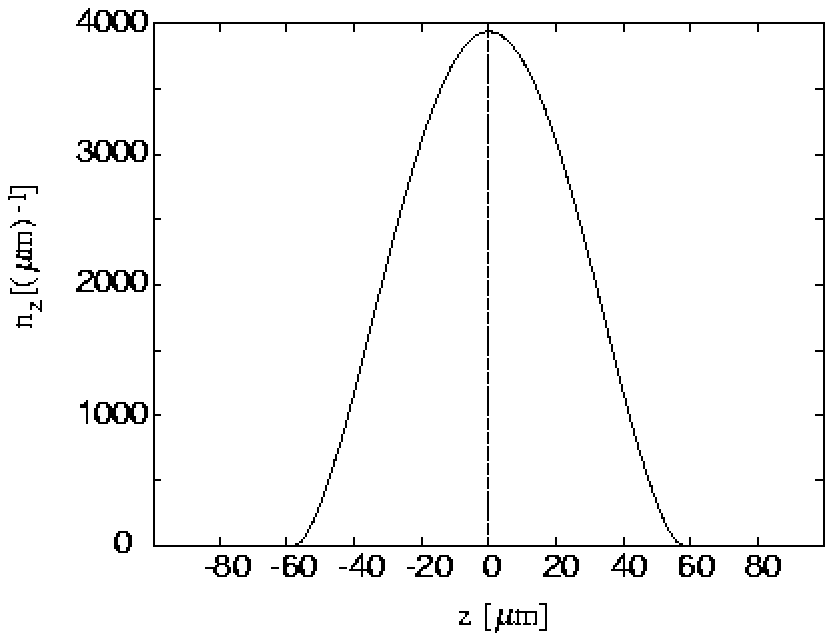}
\end{center}

(b)

Fig. 1. 
The solution of Gross-Pitaevskii equation at 
$\nu_r = 175 Hz$, $\nu_z = 10.3 Hz$.
(a) Stereographic view of the density distribution 
of the condensate $|\phi(r)|^2$.
(b) The area density $n_z$ per unit length along $z$-axis 
at the center $r=0$.
The $z$-direction of the cylindrical container is assumed 
to be uniform with the average area density, 
$n_z = 2.0 \times 10^3 / \mu m$ in the following calculation.

\begin{center}
 \leavevmode
 \epsfxsize=70mm
 \epsfbox{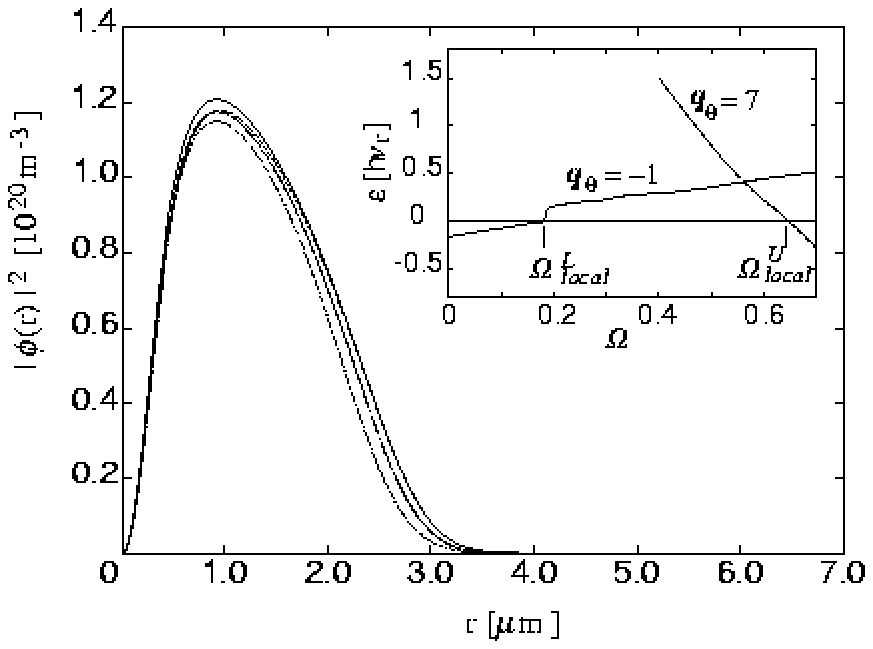}
\end{center}

(a)

\begin{center}
 \leavevmode
 \epsfxsize=70mm
 \epsfbox{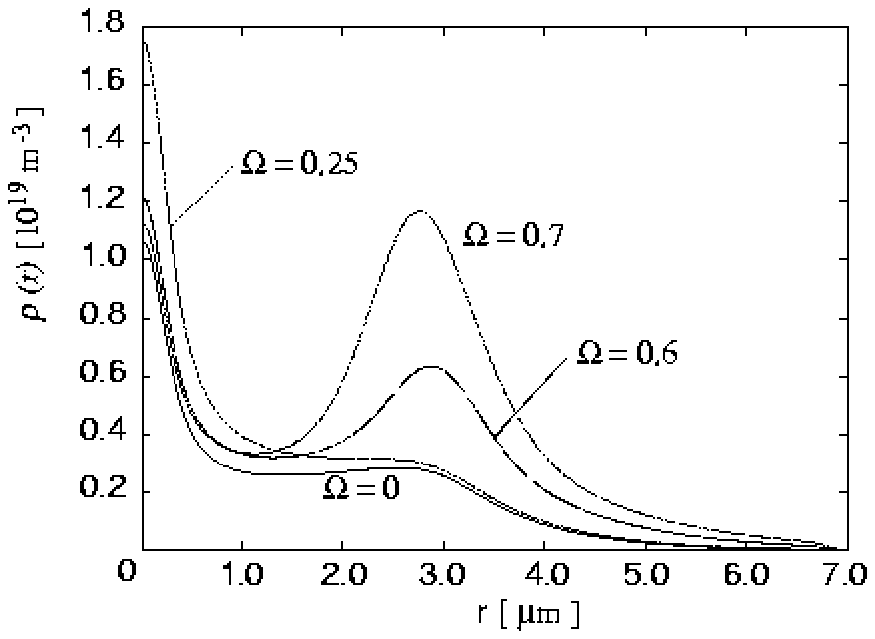}
\end{center}

(b)

\begin{center}
 \leavevmode
 \epsfxsize=70mm
 \epsfbox{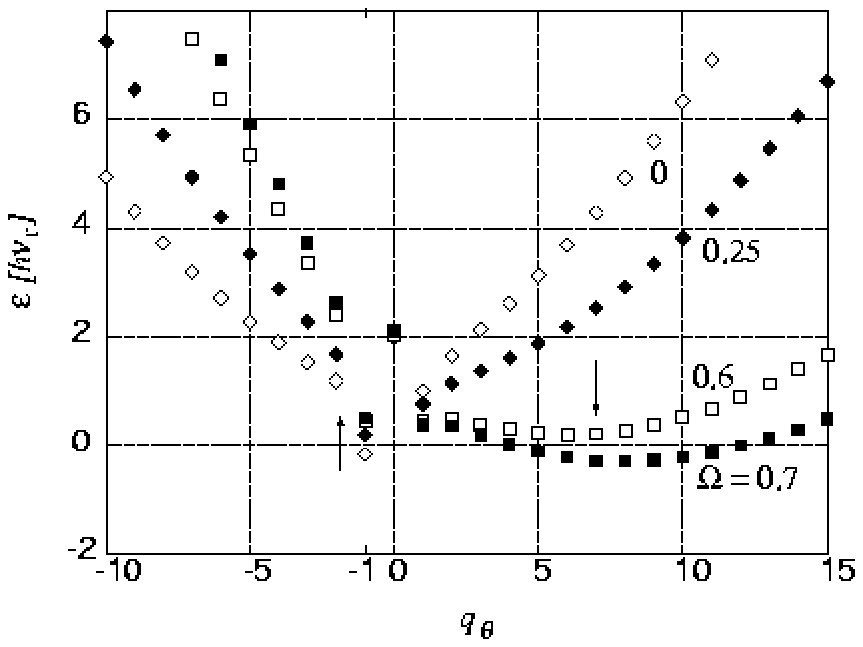}
\end{center}

(c)

Fig. 2.
At $T/T_{c0}=0.24$, 
(a) the $\Omega$ dependence of the condensate density $|\phi(r)|^2$.
The solid, dashed, dotted, dot dashed lines 
correspond to $\Omega=$ 0, 0.25, 0.6, 0.7 respectively.
(b) The $\Omega$ dependence of the non-condensate density 
$\rho(r)$.
(c) The lowest edge of the eigenvalues along $q_{\theta}$ 
for $\Omega=$ 0, 0.25, 0.6, 0.7.
The eigenvalues at $q_{\theta}=-1$ move up 
while the eigenvalues in $q_{\theta}>0$ move down as $\Omega$ increases.
The inset in (a) shows the trace of the eigenvalues at $q_{\theta}=-1$
and $q_{\theta}=7$ as a function of $\Omega$.

\begin{center}
 \leavevmode
 \epsfxsize=70mm
 \epsfbox{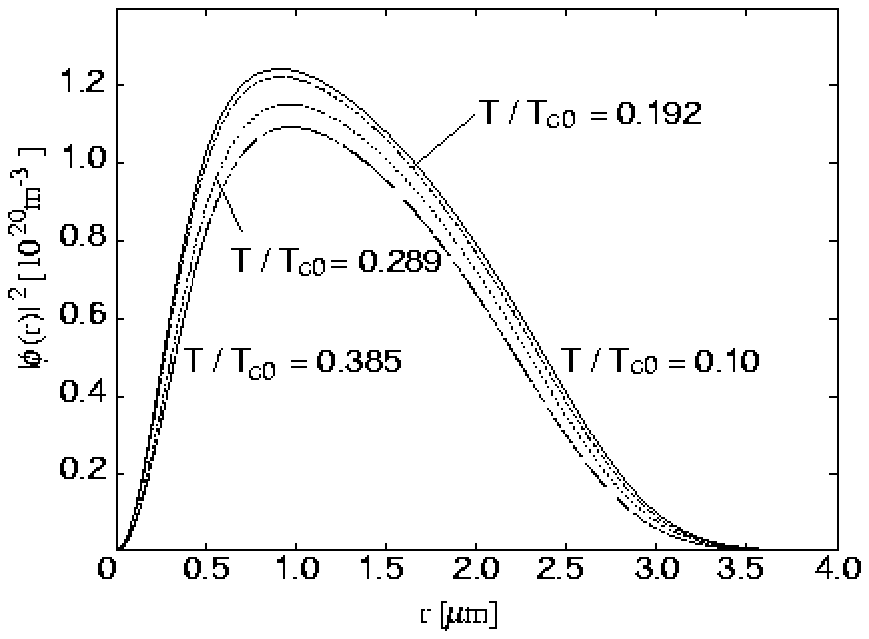}
\end{center}

(a)

\begin{center}
 \leavevmode
 \epsfxsize=70mm
 \epsfbox{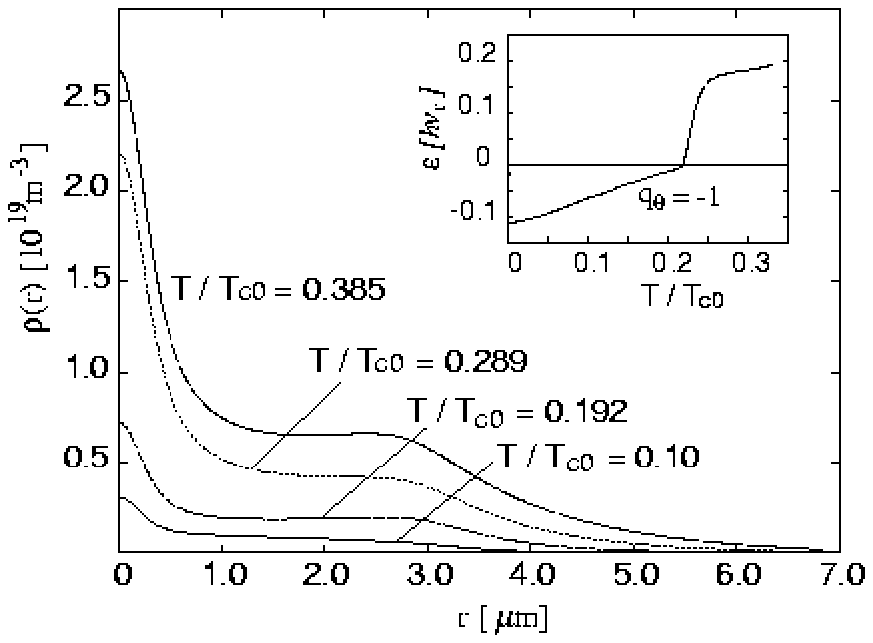}
\end{center}

(b)

Fig. 3.
At $\Omega=0.2$, 
the $T$-dependence of (a) the condensate density $|\phi(r)|^2$
and (b) the non-condensate density $\rho(r)$.
The inset of (b) shows the trace of the eigenvalue at
$q_{\theta}=-1$ as a function of $T$.

\begin{center}
 \leavevmode
 \epsfxsize=70mm
 \epsfbox{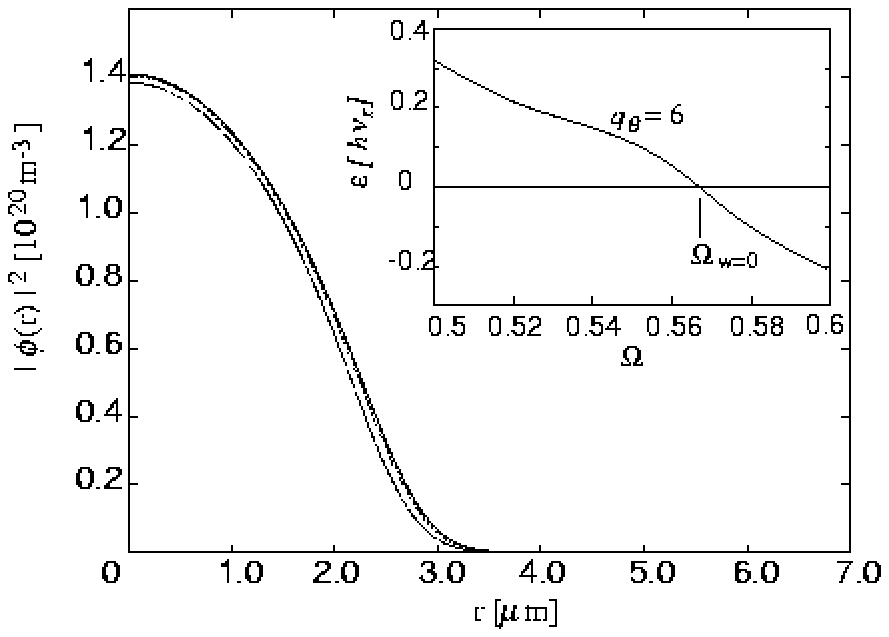}
\end{center}

(a)

\begin{center}
 \leavevmode
 \epsfxsize=70mm
 \epsfbox{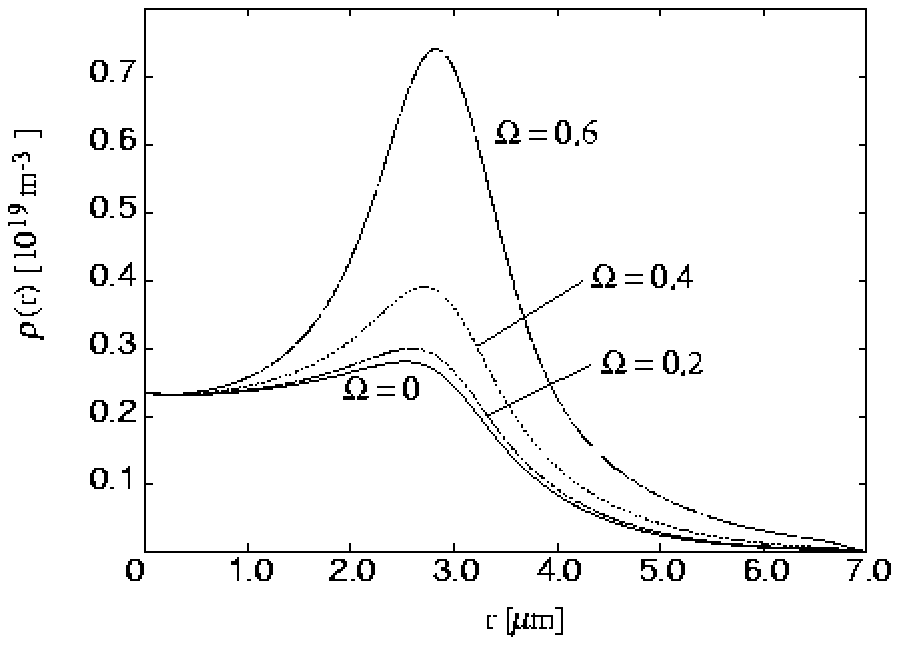}
\end{center}

(b)

\begin{center}
 \leavevmode
 \epsfxsize=70mm
 \epsfbox{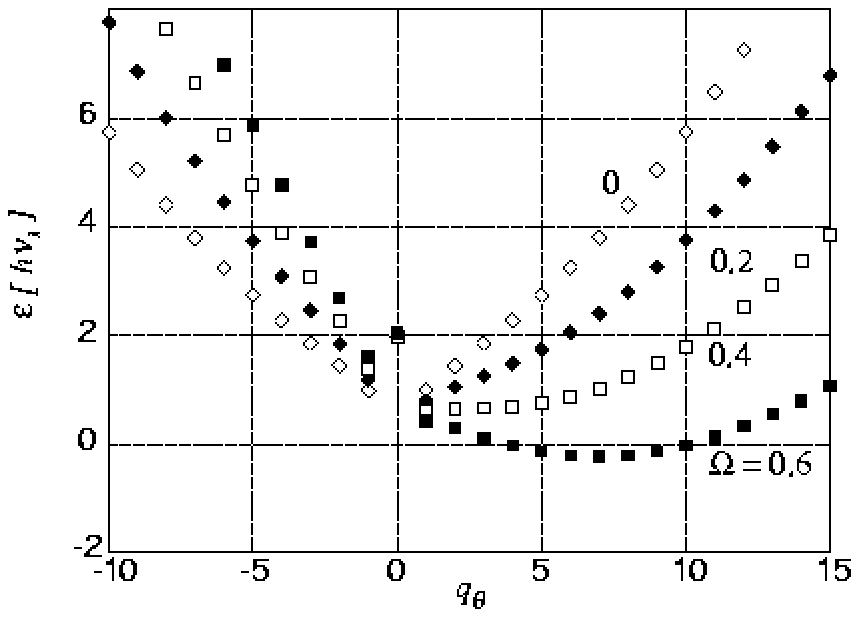}
\end{center}

(c)

Fig. 4.
At $T/T_{c0}=0.10$,
the $\Omega$-dependence of (a) the condensate density $|\phi(r)|^2$
and (b) the non-condensate density $\rho(r)$ 
for the non-vortex state.
(c) The lowest edge of the eigenvalues along $q_{\theta}$ 
for the selected $\Omega=$ 0, 0.2, 0.4, 0.6 respectively.
The positive eigenvalues in $q_{\theta}>0$ become negative 
as $\Omega$ increases.
The inset of (a) shows the trace of the eigenvalue at $q_{\theta}=6$
as a function of $\Omega$.

\begin{center}
 \leavevmode
 \epsfxsize=70mm
 \epsfbox{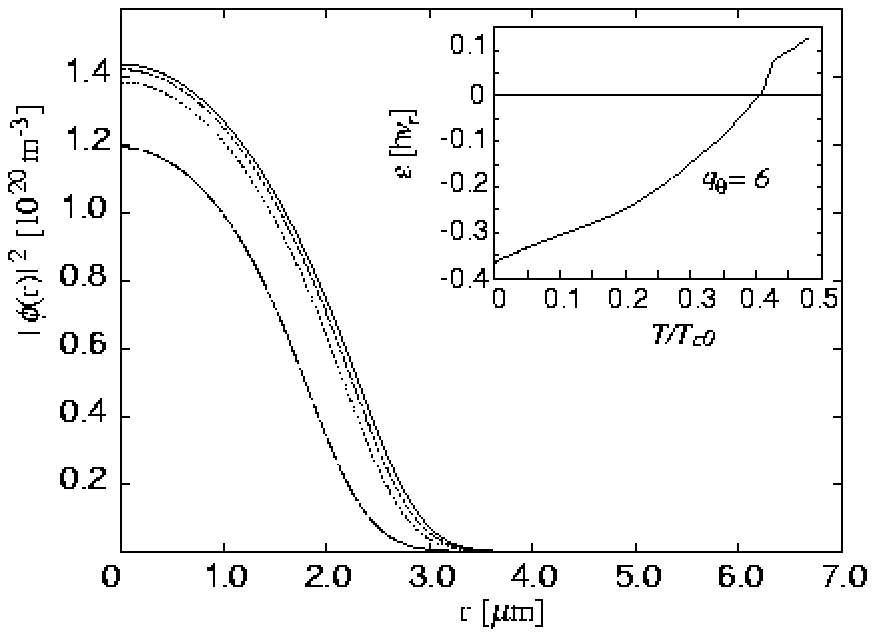}
\end{center}

(a)

\begin{center}
 \leavevmode
 \epsfxsize=70mm
 \epsfbox{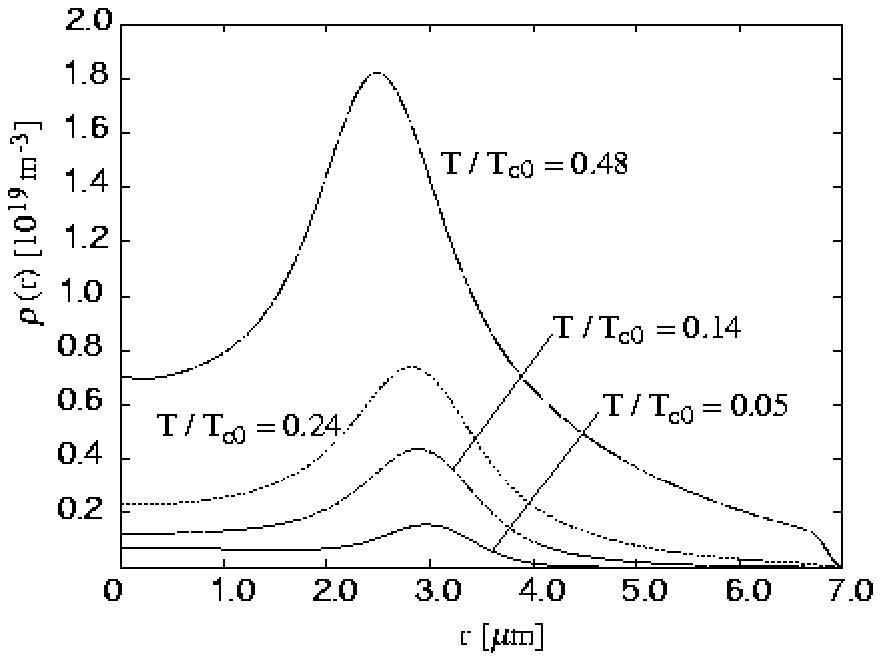}
\end{center}

(b)

Fig. 5.
At $\Omega=0.6$,
(a) the $T$-dependence of (a) the condensate density $|\phi(r)|^2$
and (b) the non-condensate density $\rho(r)$
for $T/T_{c0}=$0.05, 0.14, 0.24 and 0.48 in the non-vortex state.
The inset in (a) shows the trace of the eigenvalue at $q_{\theta}=6$
as a function of $T$.

\begin{center}
 \leavevmode
 \epsfxsize=70mm
 \epsfbox{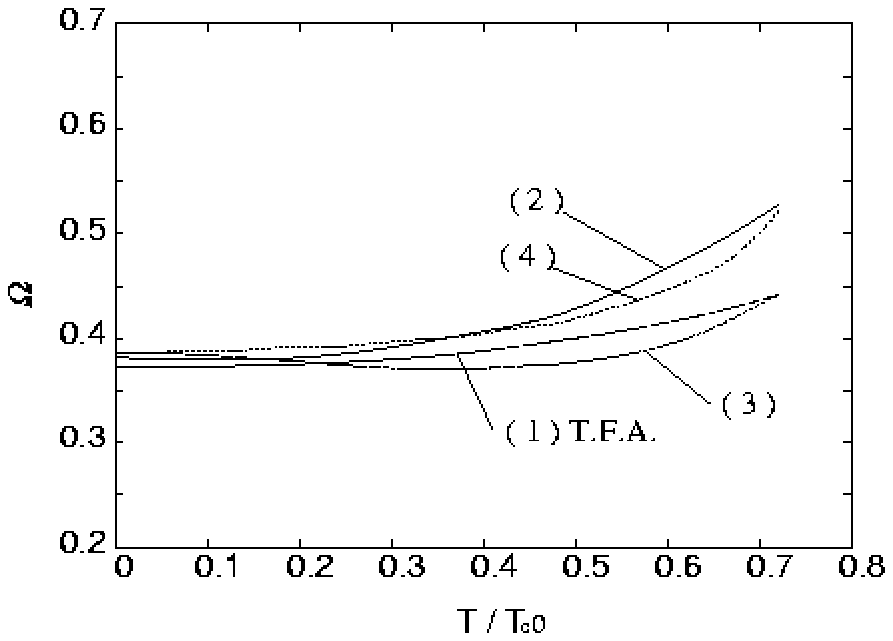}
\end{center}

Fig. 6.
Curve (1) corresponds to the critical $\Omega_{global}^L(T)$ based on
Thomas-Fermi approximation.
Curve (2) and (3) express the critical $\Omega_{global}^L(T)$ in which 
the energy of the vortex state is equal to the non-vortex state 
under same condition.
Curve (2) corresponds to $\Omega_{global}^L(T)$ estimated by neglecting 
the energy $E_0$ of the non-condensate while curve (3) corresponds to 
the critical angular velocity by comparison the total energy $E$ 
of the vortex and non-vortex state.
Curve (4) corresponds to $\Omega_{global}^L(T)$ by comparing 
the free energies $F$ of the vortex and non-vortex state.

\begin{center}
 \leavevmode
 \epsfxsize=70mm
 \epsfbox{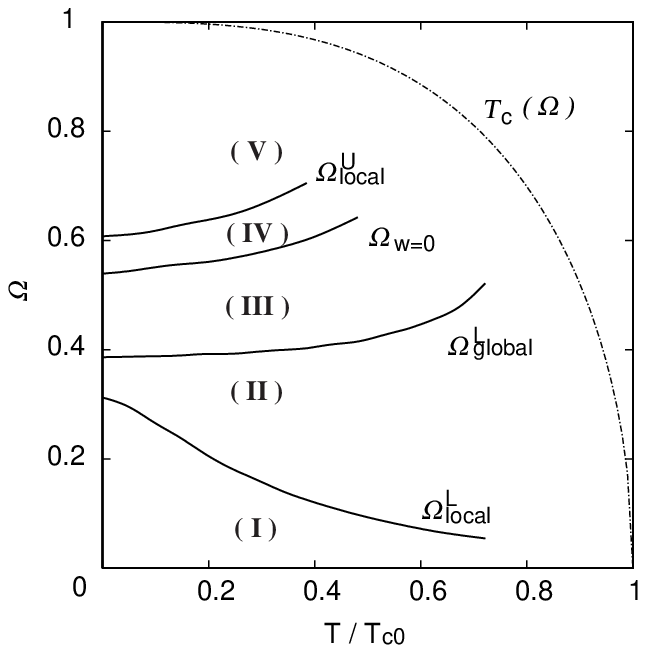}
\end{center}

(a)

\begin{center}
 \leavevmode
 \epsfxsize=80mm
 \epsfbox{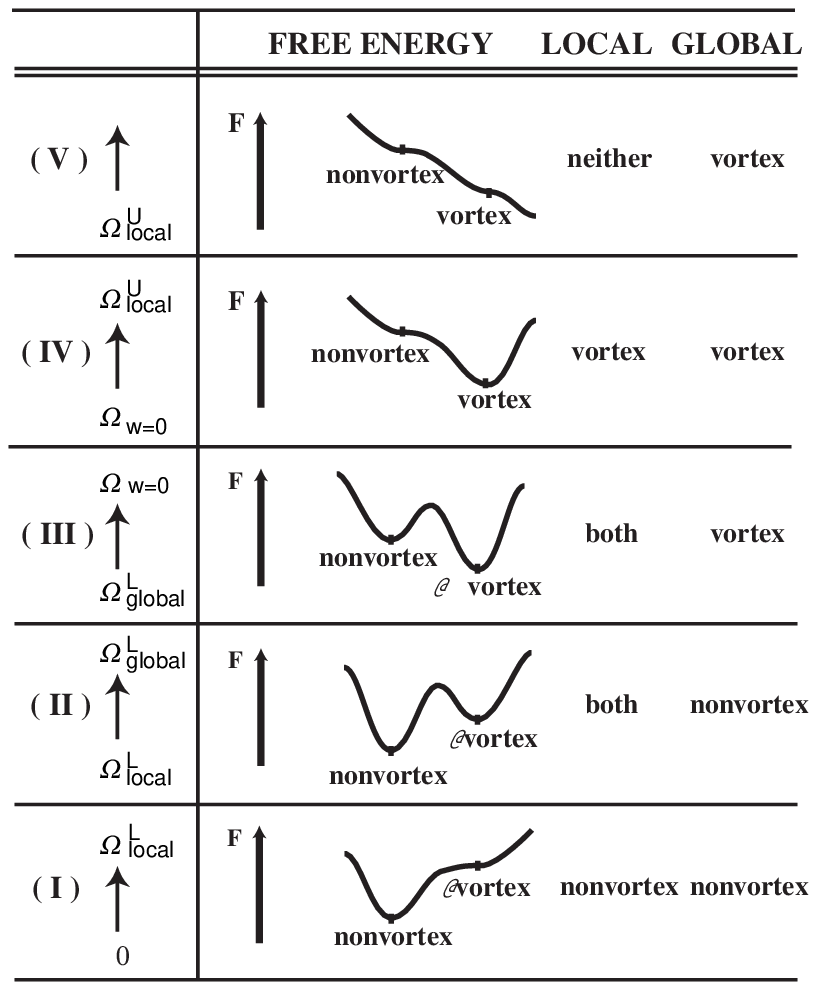}
\end{center}

(b)

Fig. 7.
Phase diagram: $\Omega$ vs $T/T_{c0}$.
(a) the four angular velocities, 
$\Omega_{local}^L$, $\Omega_{global}^L$, $\Omega_{w=0}$,
and $\Omega_{local}^U$ together with the critical temperature $T_{c}(\Omega)$
are plotted as a function of $T$.
These four lines divide the whole area into five regions, (I)-(V).
Each of the regions is explained in (b).

\begin{center}
 \leavevmode
 \epsfxsize=70mm
 \epsfbox{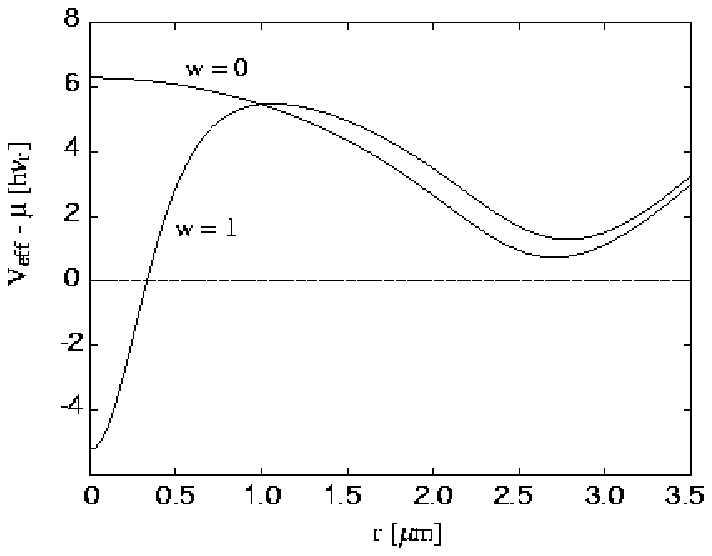}
\end{center}

(a)

\begin{center}
 \leavevmode
 \epsfxsize=70mm
 \epsfbox{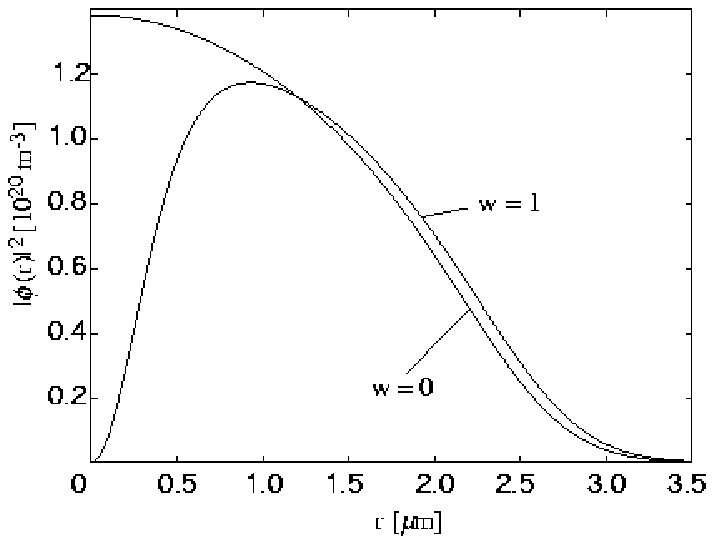}
\end{center}

(b)

Fig. 8.
At $T/T_{c0}=0.24$, $\Omega=0.6$,
(a) the effective potential $V_{eff}(r)$ 
defined by eq.(33) for the vortex state (w=1) 
and the non-vortex state (w=0)
and 
(b) the condensate density $|\phi(r)|^2$ in the vortex state
and non-vortex state. 

\end{document}